\documentclass[conference]{IEEEtran}
\IEEEoverridecommandlockouts

\usepackage{cite}
\usepackage{amsmath,amssymb,amsfonts}
\usepackage{graphicx}
\usepackage{textcomp}
\usepackage{xcolor}
\usepackage{booktabs}
\usepackage{multirow}
\usepackage{url}
\usepackage{balance}

\def\BibTeX{{\rm B\kern-.05em{\sc i\kern-.025em b}\kern-.08em
    T\kern-.1667em\lower.7ex\hbox{E}\kern-.125emX}}

\begin{document}
\bstctlcite{IEEE:BSTcontrol}

\title{From Natural Language to Silicon: The Representation Bottleneck in LLM Hardware~Design}

\author{%
Weimin~Fu$^\dag$,
Zeng~Wang$^\ddag$,
Minghao~Shao$^\ddag$$^\P$,
Johann~Knechtel$^\P$,\\
Ozgur~Sinanoglu$^\P$,
Ramesh~Karri$^\ddag$,
Muhammad~Shafique$^\P$,
Xiaolong~Guo$^\dag$,
\\
\IEEEauthorblockA{
$^\dag$Kansas State University, USA \ 
$^\ddag$NYU Tandon School of Engineering, USA \ 
$^\P$NYU Abu Dhabi, UAE\\
\normalsize{Email: \{weiminf, guoxiaolong\}}@ksu.edu}
\normalsize{\{shao.minghao, zw3464, johann, ozgursin, rkarri, muhammad.shafique\}@nyu.edu}
}

\maketitle

\begin{abstract}
Edge applications increasingly demand custom hardware, yet Field-Programmable Gate Array (FPGA) design requires expertise that domain engineers lack.
Large Language Models (LLMs) promise to bridge this gap through \emph{zero-knowledge hardware programming}, where users describe circuits in natural language and an LLM compiles them to a hardware \emph{intermediate representation} (IR) targeting silicon.
Modeling this flow as a cascade of binary filters, this work demonstrates that IR choice, not model choice, is the dominant factor governing end-to-end success, a phenomenon termed the \emph{representation bottleneck}.
An evaluation of three frontier LLMs across six IRs spanning Verilog, VHDL, Chisel, Bluespec, PyMTL3, and HLS~C on 202~tasks through a pipeline of compilation, simulation, FPGA synthesis on a Lattice iCE40UP5K, and LLM-based repair shows that simulation pass rates range from 3\% to 88\% across IRs but typically vary less than 1.25$\times$ across models within any single IR.
On the resource-constrained iCE40, LLM designs achieve a higher conditional FPGA pass rate than reference solutions, 86.5\% vs.\ 68.7\%, not because they are better but because a \emph{simplicity bias} makes them small enough to fit.
The analysis reveals an \emph{accessibility--competence paradox}: the most user-friendly IRs yield the worst LLM performance, suggesting that optimal IR selection will evolve as LLM capabilities grow.
\end{abstract}


\section{Introduction}
\label{sec:intro}

From autonomous drones to medical wearables and industrial IoT sensors, a growing class of edge applications demands low-latency, low-power compute that general-purpose embedded processors cannot efficiently deliver.
Custom hardware offers ideal performance but traditionally requires specialized design expertise.
ASICs further impose high fabrication costs and inflexible iteration cycles.
FPGAs bridge this gap: they are reprogrammable, available on low-cost development boards such as the \$10 Lattice iCE40UP5K, and deployable without foundry access.
Yet FPGA design still demands fluency in hardware description languages, making it inaccessible to the domain experts (roboticists, signal-processing engineers, embedded-systems developers) who best understand their own workloads.
With fewer than 100{,}000 hardware designers serving the global semiconductor industry~\cite{bls2025hardware}, the engineers who best understand what to compute cannot access the hardware best suited to compute it.

High-Level Synthesis (HLS) was the first systematic attempt to close this gap, allowing designers to express algorithms in C/C++ and rely on automated scheduling and allocation to produce Register-Transfer Level (RTL) netlists~\cite{cong2011hls, ferrandi2021bambu}.
HLS lowered the abstraction barrier but did not eliminate it: users must reason about memory interfaces, pipeline directives, and hardware-specific pragmas.
Large Language Models (LLMs) enable a radical step, \emph{zero-knowledge hardware programming}\footnote{The term ``zero-knowledge'' denotes ``no prior hardware design expertise,'' following~\cite{blocklove2023chipchat}, not the cryptographic notion.}~\cite{blocklove2023chipchat}, where a user with no hardware expertise describes a circuit in natural language and receives deployable code.

In this paradigm, the LLM acts as a natural language hardware compiler, and its target hardware description language such as Verilog (including SystemVerilog), VHDL, Chisel, Bluespec, PyMTL3, or HLS~C serves as the \emph{intermediate representation} (IR) of this compilation.
Because IR choice constrains both the LLM's generation quality and the downstream toolchain's ability to map the design onto silicon, it governs every stage of the flow.
This coupling is formalized as the \emph{representation bottleneck}: because the NL-to-silicon flow operates as a cascade of filters, each IR's distinct failure modes compound across stages, and no single IR dominates across all criteria.
Existing evaluations are blind to this bottleneck: all major benchmarks target Verilog~\cite{liu2023verilogeval, lu2024rtllm, thakur2023benchmarking} and stop at simulation, leaving physical implementability and IR selection unexamined.

To quantify the representation bottleneck, an evaluation pipeline is built that carries LLM-generated designs from source-language compilation through functional simulation, FPGA synthesis on a resource-constrained Lattice iCE40UP5K with 5{,}280 Look-Up Tables (LUTs), and iterative LLM-based repair.
A second target (Lattice ECP5-85K, 85\,K~LUTs) serves as a diagnostic control to separate code-quality failures from resource-limitation failures.
The evaluation covers three frontier LLMs (Claude~Sonnet~4.6, Gemini~3~Flash~Preview, GPT-5.4) across six hardware IRs on 202~design tasks drawn from VerilogEval~\cite{liu2023verilogeval} and RTLLM~\cite{lu2024rtllm}.

This paper makes the following contributions:
\begin{itemize}
    \item A cascaded-filter model of NL-to-silicon compilation in which hardware description languages serve as IRs, and the \emph{representation bottleneck} arises from the interaction of training-data availability, IR-specific lowering artifacts, and FPGA resource constraints.
    \item An end-to-end evaluation framework, the first to span multiple LLMs, multiple hardware IRs, and physical FPGA implementation, yielding 3{,}636~design evaluations across two FPGA targets.
    \item A demonstration that IR choice, not model choice, is the dominant factor: simulation pass rates vary widely across IRs (from 3\% for HLS~C to 88\% for Verilog) but typically less than 1.25$\times$ across models within any single IR. On the iCE40, LLM-generated designs achieve a higher conditional FPGA pass rate than dataset-provided reference solutions (86.5\% vs.\ 68.7\%), not because of superior design quality but because a \emph{simplicity bias} produces circuits small enough to fit the constrained fabric.
    \item An \emph{accessibility--competence paradox}: IRs most accessible to zero-knowledge users (HLS~C, PyMTL3) yield the worst LLM performance, while less accessible IRs (Chisel) achieve near-perfect FPGA pass rates, suggesting optimal IR selection will evolve with LLM capabilities.
\end{itemize}

\section{Background and Related Work}\label{sec:background}

\subsection{The Hardware IR Spectrum}

In traditional compiler design, the choice of intermediate representation determines what optimizations are possible and which targets are reachable~\cite{aho1986compilers}.
An analogous spectrum exists in hardware design: each hardware description language represents the same computation at a different level of abstraction, with distinct trade-offs in expressiveness, toolchain maturity, and training-data availability.
These three levels differ not only in abstraction but in \emph{programming model}: whether the designer \emph{describes} hardware explicitly, \emph{constructs} it through metaprogramming, or \emph{compiles} software into it.
This distinction matters because each model presents different challenges for LLM-based generation.

At the \emph{explicit RTL} level, the designer \emph{describes} hardware: signals, registers, and combinational logic are specified directly.
Verilog~\cite{ieee2005verilog} (1984) uses event-driven concurrent processes; the designer reasons in clock cycles and waveforms.
VHDL~\cite{ieee2019vhdl} (1987) takes the same approach with Ada-derived strong typing that catches errors at compile time.
Verilog dominates LLM training corpora due to its prevalence in open-source repositories~\cite{thakur2023benchmarking}; VHDL appears far less frequently.

At the \emph{hardware generation} level, the designer \emph{constructs} hardware through a host language.
Chisel~\cite{bachrach2012chisel} embeds hardware constructors in Scala: the designer writes a Scala program whose \emph{output} is a circuit, enabling parameterized generator families.
It compiles through FIRRTL to Verilog and underpins the RISC-V ecosystem, including the Rocket Chip generator~\cite{asanovic2016rocket}.
Bluespec~\cite{nikhil2004bluespec} uses guarded atomic actions where the designer specifies rules and invariants; the compiler synthesizes scheduling automatically.
Open-sourced only in 2020~\cite{nikhil2004bluespec}, its LLM training data remains scarce.
PyMTL3~\cite{jiang2020pymtl3} targets Python with a multi-level modeling methodology (functional to RTL), but remains beta software with a limited user base.

At the \emph{algorithmic} level, the designer writes software that a tool \emph{compiles} into hardware.
HLS tools such as Bambu~\cite{ferrandi2021bambu} take standard C/C++ and perform automatic scheduling, allocation, and binding to produce RTL.
The designer expresses a pure algorithm; the tool extracts parallelism.
This paradigm also introduces a different interface model: start/done handshaking and memory-bus protocols that are incompatible with standard RTL testbenches~\cite{cong2011hls}.

All non-Verilog IRs must be \emph{lowered} to Verilog through language-specific compilers before entering the FPGA synthesis flow.
This architecture directly parallels the LLVM compilation model~\cite{lattner2004llvm}: just as C, Rust, and Swift each compile to LLVM~IR before targeting x86 or ARM, Chisel, Bluespec, and HLS~C each compile to Verilog before targeting FPGA fabric.
The lowered Verilog is typically semi-structural (generated wire names, flattened hierarchies) and differs markedly from hand-written behavioral RTL.
Each lowering step also introduces independent failure points, including port renaming in Chisel's \texttt{io\_} prefixes, module wrapping in Bluespec's \texttt{mk} prefixes, and interface protocol mismatches in HLS handshaking, that are invisible to evaluations targeting Verilog alone.

\subsection{LLMs as Natural-Language-to-Hardware Compilers}

A growing body of work has demonstrated that LLMs can generate functionally correct Verilog from natural-language specifications~\cite{liu2023verilogeval, lu2024rtllm, thakur2023benchmarking, liu2025rtlcoder, pei2024betterv}.
Blocklove et al.~\cite{blocklove2023chipchat} demonstrate conversational hardware design, Chang et al.~\cite{chang2023chipgpt} explore natural-language-driven chip design, and Thakur et al.~\cite{thakur2023autochip} automate HDL generation with LLM feedback loops.
Through this IR lens, these results establish that LLMs can serve as the \emph{frontend} of a natural-language-to-silicon compiler, translating human intent into a hardware IR.

All existing evaluations, however, implicitly hard-code the IR to Verilog and evaluate only the frontend's output quality (simulation pass rate).
As frontier models approach saturation on these Verilog-only simulation benchmarks, the limitations of this single-IR evaluation paradigm become pressing.
This is analogous to evaluating a compiler solely by whether its IR type-checks, without testing whether the generated binary executes correctly on the target machine.
No prior work has studied \emph{which IR the LLM frontend should target}, nor carried the evaluation through to physical FPGA implementation.
This gap is significant because, as shown in Section~\ref{sec:results}, the choice of IR affects end-to-end success more than the choice of LLM model.

\subsection{The FPGA Backend}

The open-source FPGA ecosystem provides the backend of the compilation pipeline.
Yosys~\cite{wolf2016yosys} performs RTL synthesis with technology mapping; NextPNR handles place-and-route for Lattice iCE40 and ECP5~\cite{shah2019nextpnr}.
Unlike ASIC flows that require multi-step timing closure, FPGA synthesis yields deterministic pass/fail outcomes, making it well suited for automated evaluation.

\section{Evaluation Framework}\label{sec:methodology}

\begin{figure}[t]
\centering
\includegraphics[width=\columnwidth]{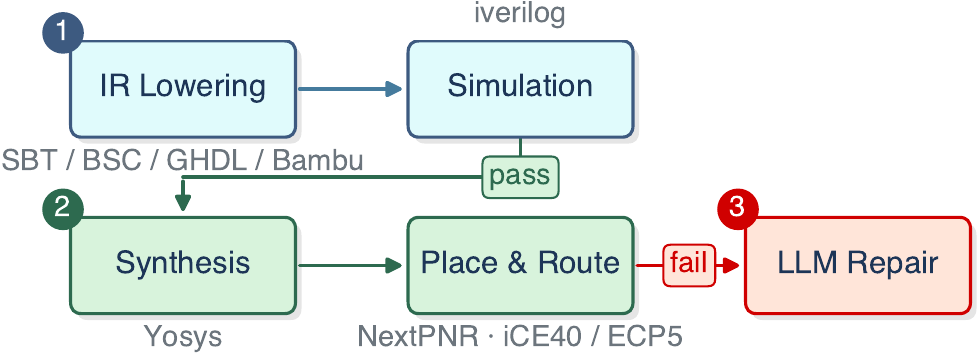}
\caption{The NL-to-silicon evaluation pipeline. Each phase isolates a distinct IR quality dimension. The orange box indicates an LLM-assisted repair step.}
\label{fig:pipeline}
\end{figure}

Studying the representation bottleneck requires a framework that evaluates the \emph{entire} NL-to-silicon compilation chain, not merely its frontend output.
The proposed pipeline (Fig.~\ref{fig:pipeline}) probes three dimensions of IR quality: (1)~\emph{lowering fidelity and functional correctness}, whether the IR can be compiled to Verilog and whether the result passes simulation; (2)~\emph{physical implementability}, whether the design can be mapped to real FPGA fabric; and (3)~\emph{repairability}, whether the LLM can recover from failures at each stage.
The pipeline targets the Lattice iCE40UP5K as the primary evaluation platform, representative of the edge deployments motivating this work.
A secondary ECP5-85K target with $\sim$17$\times$ the resources serves as a diagnostic control (Section~\ref{sec:dualtarget}).

Conceptually, this pipeline operates as a cascade of binary filters: a design must survive lowering, simulation, and synthesis in sequence.
Three properties of the cascade shape IR-level outcomes: (i)~LLM generation quality for a given IR is determined primarily by training-data availability; (ii)~each IR's lowering step introduces IR-specific, LLM-independent artifacts; and (iii)~on a resource-constrained FPGA, synthesis success is governed by design size relative to the fabric budget.
These properties generate testable predictions about IR selection that Section~\ref{sec:results} validates empirically.

\subsection{Benchmark and Models}

The benchmark comprises 202~design tasks drawn from VerilogEval~\cite{liu2023verilogeval} (156~tasks) and RTLLM~\cite{lu2024rtllm} (46~tasks), spanning combinational logic, FSMs, counters, arithmetic datapaths, and sequential circuits~\cite{fu2026synthloop}.
Each task includes a natural-language specification and a testbench.
Both benchmarks also provide reference Verilog solutions; 195 of these pass both Icarus~Verilog simulation and Yosys synthesis, forming the \emph{reference baseline} with unrestricted coding styles that may include tri-state buffers, large arrays, and complex control flow.
Three frontier LLMs are evaluated: Claude~Sonnet~4.6 (Anthropic), Gemini~3~Flash~Preview (Google), and GPT-5.4 (OpenAI), using single-attempt generation without few-shot examples to represent the baseline zero-knowledge user experience.
For non-Verilog IRs, the specification is converted into a language-specific prompt through rule-based substitution with LLM grammar correction (see Limitations) before each model generates the implementation.

\subsection{Pipeline Design}
\label{sec:pipeline}

\textbf{Phase~1: IR Lowering and Simulation.}
Each non-Verilog design is lowered to Verilog via its native compiler: GHDL~synth for VHDL, SBT/FIRRTL for Chisel, BSC for Bluespec, Python elaboration for PyMTL3, and Bambu~HLS for C.
Verilog designs pass through directly.
This phase measures IR \emph{lowering fidelity}: whether the LLM-generated code is valid in the target IR and can survive compilation.
The lowered Verilog is then simulated against the task's testbench using Icarus~Verilog.
Designs failing simulation enter up to two rounds of LLM repair at the Verilog level before proceeding. All pass rates reported in Section~\ref{sec:results} include these repair rounds.

\textbf{Phase~2: FPGA Synthesis and Place-and-Route.}
Designs passing simulation enter FPGA synthesis via Yosys and place-and-route via NextPNR, targeting the iCE40UP5K.
This phase reveals failures invisible to simulation: unsynthesizable constructs, excessive resource consumption, and routing congestion.
A design passes if NextPNR completes placement without errors.
One round of LLM repair then targets any FPGA-specific failures, such as replacing unsupported primitives or reducing resource usage, at the Verilog level with minimal-change prompting.

\subsection{Dual-Target Diagnostics}
\label{sec:dualtarget}

A design may fail FPGA implementation due to code-quality defects or resource exhaustion.
To separate these failure modes, every simulation-passing design is run through \emph{both} FPGA targets:
\begin{itemize}
    \item \textbf{iCE40UP5K} (primary): 5{,}280 logic cells, 30 Block RAMs, 96 IOs.
    \item \textbf{ECP5-85K} (diagnostic control): 83{,}640 LUTs, 3{,}744\,kbit BRAM, 365 IOs.
\end{itemize}
A design that fails on iCE40 but passes on ECP5 is resource-limited; one that fails on both has a code-quality defect.
For each design, the pipeline records pass rates at each phase per IR and per model, FPGA resource utilization and estimated $f_\text{max}$ from NextPNR, and per-design toolchain latency.
All evaluations were conducted in March 2026 using the models' default API settings, and the evaluation framework and result data will be released upon publication.

\section{Results and Analysis}\label{sec:results}

Table~\ref{tab:main} presents the full evaluation matrix: 3~LLMs $\times$ 6~IRs $\times$ 202~tasks on two FPGA targets, alongside the 195-design reference baseline.
The denominators in the Sim column vary because they reflect the number of designs that both produced valid source code and survived IR lowering to Verilog.
For Verilog itself, all 202~tasks yield compilable code.
The remaining IRs lose designs at the code-generation or lowering stage: VHDL averages 81.5\% of tasks reaching simulation, Chisel 68.3\%, Bluespec 73.6\%, PyMTL3 60.6\%, and HLS~C 98.7\%.
The high HLS~C lowering rate confirms that LLMs write valid C that Bambu can convert; the failures occur downstream.
This first-stage attrition is itself part of the representation bottleneck, and the remaining results are organized around four phenomena, each predicted by the cascaded-filter model, and one practical observation.

\begin{table}[t]
\centering
\caption{Simulation and FPGA pass rates by IR and model. Sim\% reports functional simulation pass rate after IR lowering. iCE40 and ECP5 columns report FPGA pass counts (and conditional pass rate) out of simulation-passing designs.}
\label{tab:main}
\setlength{\tabcolsep}{3pt}
\begin{tabular}{@{}llcrcc@{}}
\toprule
\textbf{IR} & \textbf{Model} & \textbf{Sim} & \textbf{Sim\%} & \textbf{iCE40} & \textbf{ECP5} \\
\midrule
\multicolumn{2}{l}{\textit{Reference baseline}} & 195/195 & 100 & 134~(68.7) & 180~(92.3) \\
\midrule
\multirow{3}{*}{Verilog}
 & Claude & 177/202 & 87.6 & 160~(90.4) & 169~(95.5) \\
 & Gemini & 167/202 & 82.7 & 141~(84.4) & 155~(92.8) \\
 & GPT    & 173/202 & 85.6 & 147~(85.0) & 167~(96.5) \\
\midrule
\multirow{3}{*}{VHDL}
 & Claude & 132/171 & 77.2 & 119~(90.2) & 129~(97.7) \\
 & Gemini & 142/182 & 78.0 & 123~(86.6) & 138~(97.2) \\
 & GPT    & 100/141 & 70.9 & 88~(88.0) & 96~(96.0) \\
\midrule
\multirow{3}{*}{Chisel}
 & Claude & 105/123 & 85.4 & 101~(96.2) & 104~(99.0) \\
 & Gemini & 103/127 & 81.1 & 90~(87.4) & 101~(98.1) \\
 & GPT    & 114/164 & 69.5 & 107~(93.9) & 114~(100) \\
\midrule
\multirow{3}{*}{Bluespec}
 & Claude & 109/165 & 66.1 & 96~(88.1) & 107~(98.2) \\
 & Gemini & 126/157 & 80.3 & 105~(83.3) & 121~(96.0) \\
 & GPT    & 87/124  & 70.2 & 78~(89.7) & 86~(98.9) \\
\midrule
\multirow{3}{*}{PyMTL3}
 & Claude & 93/127  & 73.2 & 82~(88.2) & 89~(95.7) \\
 & Gemini & 54/79   & 68.4 & 45~(83.3) & 52~(96.3) \\
 & GPT    & 112/161 & 69.6 & 99~(88.4) & 107~(95.5) \\
\midrule
\multirow{3}{*}{HLS~C}
 & Claude & 20/200  & 10.0 & 20~(100) & 13~(65.0) \\
 & Gemini & 6/196   & 3.1  & 6~(100) & 2~(33.3) \\
 & GPT    & 7/202   & 3.5  & 7~(100) & 3~(42.9) \\
\midrule
\multicolumn{2}{l}{\textbf{All LLM}} & \textbf{2{,}022} & & \textbf{1{,}748~(86.5)} & \textbf{1{,}933~(95.6)} \\
\bottomrule
\end{tabular}
\end{table}

\subsection{IR Choice Dominates Model Choice}

The central finding is that \emph{IR selection matters more than model selection}: because training-data availability and lowering behavior vary across IRs rather than across models, the cascaded-filter model predicts this pattern.
Within a single IR such as Verilog, the three models span a 6~percentage-point (pp) range on iCE40 (84\%--90\%).
Across IRs for a single model such as Claude, iCE40 pass rates span 12\,pp (88\%--100\%).
Simulation pass rates show the effect more directly: Verilog achieves 83\%--88\% across models, while HLS~C achieves only 3\%--10\%, a gap driven largely by the HLS toolchain's interface mismatch as discussed below.
Holding the IR fixed and varying the model produces less than a 1.25$\times$ difference for all established IRs; HLS~C shows a larger ratio (3.2$\times$), but its small sample size (6--20 simulation-passing designs) limits the significance of this outlier.
This pattern mirrors the LLVM ecosystem~\cite{lattner2004llvm}, where the choice of source language affects the quality of generated IR more than the choice of optimization level, and it suggests that improving LLM hardware generation should prioritize \emph{IR-aware training} rather than model scaling alone.
A per-task analysis confirms that IR failure modes are largely independent: of the 202 tasks, only 17 (8\%) pass simulation in all six IRs and only 10 (5\%) fail in all six, while the remaining 175 (87\%) show IR-dependent success patterns where different IRs fail on different tasks.

\subsection{The Simulation Bottleneck}

Because iCE40 synthesis is primarily a resource-threshold test, designs surviving the harder simulation filter should map to silicon at high rates.
Indeed, across all IRs, once a design passes simulation, iCE40 FPGA pass rates are uniformly high (83\%--100\%).
The representation bottleneck therefore operates at the \emph{compilation and simulation} layer, not at synthesis, meaning the bottleneck is a \emph{frontend} problem.
The cascade structure is visible in the raw attrition rates: the first filter (IR lowering) retains 100\% of Verilog designs but only 61\% of PyMTL3 and 68\% of Chisel on average; the second filter (simulation) retains 70\%--85\% of lowered designs depending on the IR; the third filter (FPGA synthesis) retains 83\%--100\% of simulation-passing designs regardless of IR.
Each stage compounds the previous attrition, but the third filter is nearly transparent, confirming that the frontend dominates.
Simulation pass rates reveal each IR's frontend quality directly.
Verilog (83\%--88\%) benefits from the largest training corpus~\cite{thakur2023benchmarking}.
Chisel (70\%--85\%) and VHDL (71\%--78\%) impose an additional lowering step but produce largely compilable code.
Bluespec (66\%--80\%) suffers from scarce training data, though its guarded-atomic-action model~\cite{nikhil2004bluespec} yields structurally clean designs once compiled.
PyMTL3 (68\%--73\%) is hindered by its beta-quality framework and limited documentation~\cite{jiang2020pymtl3}.
HLS~C (3\%--10\%) fails at a fundamentally different level, and the nature of its failure is instructive.
The LLMs \emph{can} write valid C: Bambu successfully converts the C code to Verilog in the vast majority of cases.
The failure occurs downstream, in the mismatch between the HLS-generated Verilog interface (start/done handshaking, memory-bus protocols) and the standard RTL testbenches.
This is not an LLM competence problem but a \emph{toolchain ecosystem} problem: the IR's lowering step produces Verilog that is structurally incompatible with the verification infrastructure.
The HLS~C anomaly in Fig.~\ref{fig:dualtarget} reinforces this interpretation: the few HLS designs that survive simulation achieve 100\% iCE40 pass rate but only 47\% on ECP5, because the two synthesis backends handle HLS-generated Verilog patterns differently.
Teaching LLMs the idioms and constraints of each IR will yield larger gains than improving the synthesis backend.

The bottleneck is not uniform across design categories~\cite{fu2026synthloop}: combinational logic shows a 1.3--1.4$\times$ IR gap, while stateful designs (counters, sequences) widen it to 2.3$\times$, and memory designs are the hardest across all IRs (43\% aggregate).
The representation bottleneck widens as designs require more complex state management, because sequential semantics differ more across IRs than combinational logic does.

Simulation repair (up to two rounds) partially mitigates frontend failures, but its effectiveness varies by IR: repair rescues 41\% of failing PyMTL3 designs and 33\% of Verilog designs, but only 10\%--22\% for VHDL, Chisel, and Bluespec, and under 2\% for HLS~C.
The repair gradient broadly tracks training-data availability, with the exception of PyMTL3, whose simpler error patterns appear more amenable to iterative correction.

\subsection{Simplicity Bias and the Reference Baseline}

LLMs trained predominantly on common code patterns produce structurally simple designs, and the cascade's survivorship filtering further selects for small circuits.
On the resource-constrained iCE40, this simplicity becomes an advantage: LLM-generated designs achieve an aggregate 86.5\% FPGA pass rate, while the dataset-provided reference solutions achieve only 68.7\%.
Note that these are conditional FPGA pass rates (given simulation success). The unconditional end-to-end success rate for LLM designs is 1{,}748/3{,}636 = 48.1\%, reflecting the cumulative cost of the representation bottleneck across all pipeline stages. For Verilog alone, the unconditional rate is 74.0\%, comparable to the reference baseline's 68.7\%.
This result requires careful interpretation.

\begin{figure}[t]
\centering
\includegraphics[width=\columnwidth]{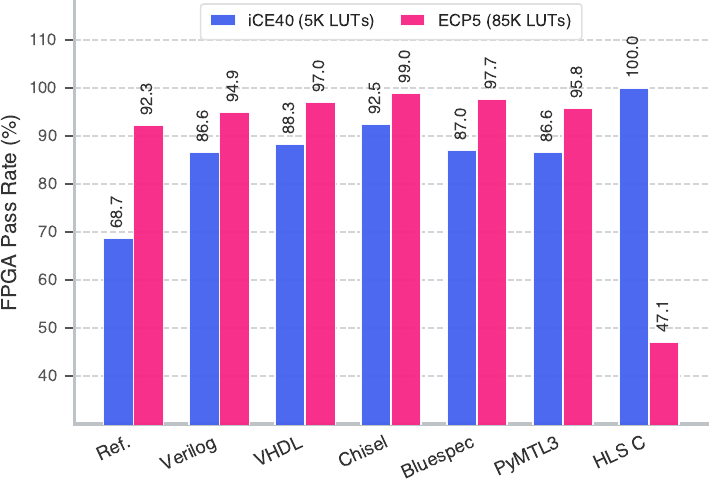}
\caption{FPGA pass rates by IR (averaged across three LLMs) on iCE40 vs.\ ECP5. The reference baseline gains 23.6\,pp on the larger device, confirming that its iCE40 failures are resource-related. LLM designs gain only 9.1\,pp because their simpler implementations already fit within iCE40's constraints.}
\label{fig:dualtarget}
\end{figure}

The reference solutions are correct implementations provided by the benchmark datasets.
They may use advanced Verilog features (tri-state buffers, large arrays, complex control flow) that produce designs too large for iCE40's 5{,}280 logic cells.
The dual-target diagnostic (Fig.~\ref{fig:dualtarget}) confirms this: the reference baseline gains 23.6\,pp moving from iCE40 to ECP5 (68.7\% $\rightarrow$ 92.3\%), showing that most of its iCE40 failures are resource-related.
LLM designs gain only 9.1\,pp (86.5\% $\rightarrow$ 95.6\%), because they already fit within iCE40's constraints.

Table~\ref{tab:resources} quantifies the underlying mechanism: non-Verilog IRs produce notably smaller circuits (average LCs: VHDL~13, Bluespec~11, PyMTL3~10) compared to Verilog (24) and the reference (20).
A survivorship effect amplifies this: designs that fail simulation never enter the FPGA pipeline, further skewing toward simpler circuits.
Because different subsets of the 202 tasks survive for each IR (N ranges from 33 to 448), the per-IR statistics characterize each IR's surviving population rather than a controlled comparison.

\begin{table}[t]
\centering
\caption{FPGA resource utilization and toolchain cost on iCE40UP5K.}
\label{tab:resources}
\begin{tabular}{@{}lrrrrc@{}}
\toprule
\textbf{IR} & \textbf{N} & \textbf{LC avg} & \textbf{LC max} & \textbf{$f_\text{max}$} & \textbf{Time} \\
 & & & & \textbf{(MHz)} & \textbf{(s/design)} \\
\midrule
Reference & 134 & 20 &  916 & 167 & 0.1 \\
Verilog   & 448 & 24 & 1{,}072 & 162 & 0.1 \\
VHDL      & 330 & 13 &  107 & 164 & 10.8 \\
Chisel    & 298 & 20 & 2{,}150 & 165 & 29.5 \\
Bluespec  & 279 & 11 &  108 & 164 & 11.5 \\
PyMTL3    & 226 & 10 &  190 & 174 & 0.2 \\
HLS~C     &  33 &  2 &    2 & N/A & 18.2 \\
\bottomrule
\end{tabular}
\end{table}

The simplicity bias is not a sign of superior design quality but a consequence of how LLMs generate code; for resource-constrained FPGAs where budgets are strict, the net effect is positive because LLM designs fit within the constraints that reference solutions exceed.

\subsection{The Accessibility--Competence Paradox}

Higher-abstraction IRs should, in principle, benefit zero-knowledge users twice: they are easier for humans to understand \emph{and} their structural regularity improves synthesis quality (Chisel achieves 92.5\% iCE40 pass rate for simulation-passing designs).
Yet the data show the opposite for generation quality: the IRs most \emph{accessible} to zero-knowledge users, HLS~C and PyMTL3, are those where LLMs perform worst (3\%--10\% and 68\%--73\% simulation pass rates), while Verilog, the IR most demanding of hardware expertise, yields the best generation quality (83\%--88\%).
The root cause is that LLM competence tracks \emph{training data availability}, not \emph{language abstraction level}: Verilog dominates open-source hardware repositories, while Bambu HLS examples are rare and PyMTL3's user base is limited to a single research group.
The paradox is thus a tension between two desirable properties, synthesis friendliness and generation quality, that are currently anticorrelated due to training-data economics.

This tension is not permanent, because training data is a function of time and investment.
As LLM capabilities expand to cover minority IRs through targeted fine-tuning or synthetic data generation, the optimal IR choice will shift toward higher-abstraction IRs whose structural advantages in synthesis can then be realized.
This evolution is discussed in Section~\ref{sec:discussion}.

\subsection{Toolchain Cost}

Practical IR selection must also consider toolchain latency.
Table~\ref{tab:resources} shows a 300$\times$ spread: Verilog completes in $\sim$0.1\,s per design while Chisel requires $\sim$30\,s due to JVM startup and Scala compilation, a significant cost for interactive workflows.
At the same time, $f_\text{max}$ is consistent across IRs (162--174~MHz on iCE40), indicating that the synthesis backend normalizes structural differences introduced by different IRs.
The choice of IR therefore affects \emph{whether} a design reaches the FPGA, not \emph{how well} it performs once there.

\section{Discussion: The Evolution of IR Selection}\label{sec:discussion}

The representation bottleneck as characterized in Section~\ref{sec:results} is not static; as LLMs improve, the optimal IR will evolve.

\textbf{Training data will shift the frontier.}
Today, Verilog leads because it dominates LLM training corpora, a contingent advantage.
Targeted fine-tuning and synthetic data generation will progressively close the gap for higher-abstraction IRs.
The question is not \emph{whether} LLMs will become proficient in Chisel or Bluespec, but \emph{when}.

\textbf{Higher-abstraction IRs have latent advantages.}
Chisel already achieves 99\% FPGA pass rates on ECP5 despite moderate simulation rates, because its hardware-constructor model~\cite{bachrach2012chisel} produces structurally regular, synthesis-friendly Verilog.
Bluespec's guarded-atomic-action semantics~\cite{nikhil2004bluespec} could offer even stronger guarantees: the compiler handles scheduling automatically, eliminating an entire class of concurrency bugs.
As LLM frontend quality improves for these IRs, their structural advantages in the backend (lowering and synthesis) will compound.
The data in Table~\ref{tab:main} supports this: among all IRs, Chisel shows the highest iCE40 pass rate (92.5\%) for designs that survive simulation, suggesting that its lowered Verilog is inherently more synthesis-friendly.

\textbf{The IR spectrum may itself evolve.}
The finding that the representation layer, not the model, is the binding constraint aligns with a growing recognition across the hardware design community that abstraction layers constitute the most fundamental bottleneck in scaling chip design to meet accelerating demand.
The current IR landscape was designed for human programmers, and an LLM-native IR that explicitly separates \emph{intent} from \emph{implementation} could further reduce the representation bottleneck.
The proposed framework provides the evaluation infrastructure to assess such future IRs on equal footing with existing ones.
At present, Verilog remains the pragmatic default for zero-knowledge users who need reliable generation.
For users who can tolerate higher toolchain latency, Chisel offers the best combination of simulation pass rate and FPGA synthesis reliability, and its advantages will grow as LLM training data for Scala-embedded hardware expands.
HLS~C, despite its conceptual appeal for non-hardware programmers~\cite{cong2011hls}, is not viable for LLM-based generation in its current form due to the interface protocol mismatch and the scarcity of synthesizable C training examples.

\subsection{Limitations}

\textbf{Benchmark and evaluation scope.}
VerilogEval and RTLLM contain educational-scale designs with a median of fewer than 50 lines.
Industrial designs with complex hierarchies and IP integration may shift the IR ranking.
All results use single-attempt zero-shot prompting, which directly models the zero-knowledge user scenario: providing few-shot examples presupposes the hardware expertise the paradigm aims to eliminate.
More sophisticated prompting or agentic flows may change absolute pass rates, though the \emph{relative} IR ranking is likely stable because the underlying training-data imbalance persists across prompting strategies.
The study does not include confidence intervals or repeated runs; while 202 design tasks provide reasonable statistical power in aggregate, individual per-IR results for small populations such as HLS~C should be interpreted with caution.

\textbf{HLS interface mismatch.}
The low HLS~C simulation pass rate partly reflects the mismatch between HLS-generated interfaces and standard RTL testbenches, not solely LLM generation quality.
HLS-native verification could improve these results, but the interface mismatch is itself a manifestation of the representation bottleneck: an IR whose lowered output requires a completely different verification methodology imposes additional cost on the end-to-end flow.

\textbf{Prompt conversion for non-Verilog IRs.}
For non-Verilog IRs, the Verilog-oriented specification is converted to a language-specific prompt through rule-based substitution (replacing language names, keywords, and module conventions), with Gemini~3~Pro~Preview performing only English grammar correction.
Because the conversion is predominantly mechanical rather than generative, it does not introduce an LLM-dependent confound.
All designs are verified against the \emph{same} Verilog testbenches regardless of source IR, and the conversion is applied uniformly across all three evaluation models.

\textbf{FPGA targets.}
Results are specific to Lattice iCE40 and ECP5 via open-source tools.
Commercial Xilinx/Intel flows may yield different absolute numbers, though the relative IR effects should persist since they originate in the frontend, not in the backend synthesis tool.

\section{Conclusion}\label{sec:conclusion}

This work has framed hardware description languages as intermediate representations in the natural-language-to-silicon compilation performed by LLMs, and shown that IR selection, not model selection, is the dominant factor governing end-to-end success.
Through an end-to-end evaluation framework spanning 3~LLMs, 6~hardware IRs, 202~tasks, and 2~FPGA targets, the analysis identifies three structural phenomena:
the \emph{simulation bottleneck}, where compilation and simulation, not FPGA synthesis, are the primary filter;
the \emph{simplicity bias}, where LLM designs outperform reference solutions on the iCE40 because they produce smaller circuits, not better ones; and
the \emph{accessibility--competence paradox}, where the most user-friendly IRs yield the worst LLM performance due to training-data scarcity.
These findings carry a practical message for zero-knowledge hardware designers: \emph{choose your IR carefully}.
Today, Verilog is the pragmatic choice; Chisel offers the best balance of generation quality and synthesis reliability for users who can tolerate its toolchain latency.
As LLMs become proficient in higher-abstraction IRs, the training-data component of the bottleneck will close, but the architectural components, the cascade of independent filters and IR-specific lowering artifacts, will persist, ensuring that different IRs will always present distinct trade-offs in expressiveness, verifiability, and synthesis efficiency.
Understanding these trade-offs is essential for building the next generation of NL-to-silicon tools.

\balance
\bibliographystyle{IEEEtran}
\bibliography{ref}

@IEEEtranBSTCTL{IEEE:BSTcontrol,
  CTLuse_forced_etal       = "yes",
  CTLmax_names_forced_etal = "3",
  CTLnames_show_etal       = "2"
}

@misc{bls2025hardware,
  author       = {{U.S. Bureau of Labor Statistics}},
  title        = {Occupational Outlook Handbook: Computer Hardware Engineers},
  year         = {2025},
  howpublished = {\url{https://www.bls.gov/ooh/architecture-and-engineering/computer-hardware-engineers.htm}},
  note         = {76{,}800 jobs in 2024; accessed March 2026}
}

@misc{fu2026synthloop,
  title={Synthesis-in-the-Loop Evaluation of {LLMs} for {RTL} Generation: Quality, Reliability, and Failure Modes},
  author={Fu, Weimin and Wang, Zeng and Shao, Minghao and Karri, Ramesh and Shafique, Muhammad and Knechtel, Johann and Sinanoglu, Ozgur and Guo, Xiaolong},
  year={2026},
  eprint={2603.11287},
  archivePrefix={arXiv},
  primaryClass={cs.AR}
}

@inproceedings{liu2023verilogeval,
  title={{VerilogEval}: Evaluating Large Language Models for {Verilog} Code Generation},
  author={Liu, Mingjie and Pinckney, Nathaniel and Khailany, Brucek and Ren, Haoxing},
  booktitle={Proc. IEEE/ACM Int. Conf. Computer-Aided Design (ICCAD)},
  year={2023}
}

@inproceedings{lu2024rtllm,
  title={{RTLLM}: An Open-Source Benchmark for Design {RTL} Generation with Large Language Model},
  author={Lu, Yao and Liu, Shang and Zhang, Qijun and Xie, Zhiyao},
  booktitle={Proc. Asia and South Pacific Design Automation Conf. (ASP-DAC)},
  year={2024}
}

@inproceedings{thakur2023benchmarking,
  title={Benchmarking Large Language Models for Automated {Verilog RTL} Code Generation},
  author={Thakur, Shailja and Ahmad, Baleegh and Fan, Zhenxing and Pearce, Hammond and Tan, Benjamin and Karri, Ramesh and Dolan-Gavitt, Brendan and Garg, Siddharth},
  booktitle={Proc. Design, Automation and Test in Europe (DATE)},
  year={2023}
}

@inproceedings{blocklove2023chipchat,
  title={{Chip-Chat}: Challenges and Opportunities in Conversational Hardware Design},
  author={Blocklove, Jason and Garg, Siddharth and Karri, Ramesh and Pearce, Hammond},
  booktitle={Proc. IEEE/ACM Int. Conf. on Machine Learning for EDA (MLCAD)},
  year={2023}
}

@inproceedings{bachrach2012chisel,
  title={Chisel: Constructing Hardware in a {Scala} Embedded Language},
  author={Bachrach, Jonathan and Vo, Huy and Richards, Brian and Lee, Yunsup and Waterman, Andrew and Avi{\v{z}}ienis, Rimas and Wawrzynek, John and Asanovi{\'c}, Krste},
  booktitle={Proc. ACM/IEEE Design Automation Conf. (DAC)},
  year={2012}
}

@article{jiang2020pymtl3,
  title={{PyMTL3}: A {Python} Framework for Open-Source Hardware Modeling, Generation, Simulation, and Verification},
  author={Jiang, Shunning and Pan, Peitian and Ou, Yanghui and Batten, Christopher},
  journal={IEEE Micro},
  volume={40},
  number={4},
  pages={58--66},
  year={2020}
}

@inproceedings{nikhil2004bluespec,
  title={{Bluespec System Verilog}: Efficient, Correct {RTL} from High Level Specifications},
  author={Nikhil, Rishiyur S.},
  booktitle={Proc. ACM/IEEE Int. Conf. Formal Methods and Models for Co-Design (MEMOCODE)},
  year={2004}
}

@inproceedings{ferrandi2021bambu,
  title={Bambu: An Open-Source Research Framework for the High-Level Synthesis of Complex Applications},
  author={Ferrandi, Fabrizio and Ferro, Andrea and Castellana, Vito Giovanni and Curzel, Serena and Fezzardi, Pietro and Fiorito, Michele and Galimberti, Marco and Lattuada, Marco and Minutoli, Marco and Pilato, Christian and Tumeo, Antonino},
  booktitle={Proc. ACM/IEEE Design Automation Conf. (DAC)},
  year={2021}
}

@inproceedings{wolf2016yosys,
  title={Yosys -- A Free {Verilog} Synthesis Suite},
  author={Wolf, Clifford},
  booktitle={Proc. Austrochip Workshop on Microelectronics},
  year={2016}
}

@book{aho1986compilers,
  title={Compilers: Principles, Techniques, and Tools},
  author={Aho, Alfred V. and Sethi, Ravi and Ullman, Jeffrey D.},
  publisher={Addison-Wesley},
  year={1986}
}

@article{cong2011hls,
  title={High-Level Synthesis for {FPGAs}: From Prototyping to Deployment},
  author={Cong, Jason and Liu, Bin and Neuendorffer, Stephen and Noguera, Juanjo and Vissers, Kees A. and Zhang, Zhiru},
  journal={IEEE Trans. Computer-Aided Design of Integrated Circuits and Systems},
  volume={30},
  number={4},
  pages={473--491},
  year={2011}
}

@inproceedings{lattner2004llvm,
  title={{LLVM}: A Compilation Framework for Lifelong Program Analysis \& Transformation},
  author={Lattner, Chris and Adve, Vikram},
  booktitle={Proc. IEEE/ACM Int. Symp. Code Generation and Optimization (CGO)},
  year={2004}
}

@inproceedings{shah2019nextpnr,
  title={Yosys+nextpnr: An Open Source Framework from {Verilog} to Bitstream for Commercial {FPGAs}},
  author={Shah, David and Hung, Eddie and Wolf, Clifford and Bazanski, Serge and Gisselquist, Dan and Milanovic, Miodrag},
  booktitle={Proc. IEEE Int. Symp. Field-Programmable Custom Computing Machines (FCCM)},
  pages={1--4},
  year={2019}
}

@techreport{asanovic2016rocket,
  title={The {Rocket Chip} Generator},
  author={Asanovi{\'c}, Krste and Avi{\v{z}}ienis, Rimas and Bachrach, Jonathan and Beamer, Scott and Biancolin, David and Celio, Christopher and Cook, Henry and Dabbelt, Daniel and Hauser, John and Izraelevitz, Adam and Karandikar, Sagar and Keller, Ben and Kim, Donggyu and Koenig, John and Lee, Yunsup and Love, Eric and Maas, Martin and Magyar, Albert and Mao, Howard and Moreto, Miquel and Ou, Albert and Patterson, David A. and Richards, Brian and Schmidt, Colin and Twigg, Stephen and Vo, Huy and Waterman, Andrew},
  institution={EECS Department, University of California, Berkeley},
  number={UCB/EECS-2016-17},
  year={2016}
}

@standard{ieee2005verilog,
  title={{IEEE} Standard for {Verilog} Hardware Description Language},
  organization={IEEE},
  number={1364-2005},
  year={2005}
}

@standard{ieee2019vhdl,
  title={{IEEE} Standard for {VHDL} Language Reference Manual},
  organization={IEEE},
  number={1076-2019},
  year={2019}
}

@article{chang2023chipgpt,
  title={{ChipGPT}: How Far Are We From Natural Language Hardware Design},
  author={Chang, Kaiyan and Wang, Ying and Ren, Haimeng and Wang, Mengdi and Liang, Shengwen and Han, Yinhe and Li, Huawei and Li, Xiaowei},
  journal={arXiv preprint arXiv:2305.14019},
  year={2023}
}

@article{liu2025rtlcoder,
  title={{RTLCoder}: Fully Open-Source and Efficient {LLM}-Assisted {RTL} Code Generation Technique},
  author={Liu, Shang and Fang, Wenji and Lu, Yao and Wang, Jing and Zhang, Qijun and Zhang, Hongce and Xie, Zhiyao},
  journal={IEEE Trans. Computer-Aided Design of Integrated Circuits and Systems},
  volume={44},
  number={4},
  pages={1448--1461},
  year={2025}
}

@article{thakur2023autochip,
  title={{AutoChip}: Automating {HDL} Generation Using {LLM} Feedback},
  author={Thakur, Shailja and Blocklove, Jason and Pearce, Hammond and Tan, Benjamin and Garg, Siddharth and Karri, Ramesh},
  journal={arXiv preprint arXiv:2311.04887},
  year={2023}
}

@inproceedings{pei2024betterv,
  title={{BetterV}: Controlled {Verilog} Generation with Discriminative Guidance},
  author={Pei, Zehua and Zhen, Hui-Ling and Yuan, Mingxuan and Huang, Yu and Yu, Bei},
  booktitle={Proc. Int. Conf. Machine Learning (ICML)},
  year={2024}
}

\end{document}